\documentclass{article}

\usepackage{epsfig}
\begin{document}
\begin{center}
\noindent \\[0pt]
{\bf\Large Are non-magnetic mechanisms such as temporal solar diameter variations conceivable for an irradiance variability?}\vspace*{1cm}
\end{center}

\begin{center}
\noindent \\[0pt]
{\Large J.P. Rozelot}\\[0pt]
Observatoire de la C\^{o}te d'Azur, D\'epartement GEMINI, Grasse, France\\[0pt]
jean-pierre.rozelot@obs-azur.fr \\
\vspace*{0.25cm}
{\Large S. Lefebvre}\\[0pt]
Department of Physics and Astronomy, UCLA, Los Angeles, USA\\[0pt]
\vspace*{0.25cm}
{\Large S. Pireaux}\\[0pt]
Observatoire Midi-Pyr\'en\'ees, UMR 5562-DTP, Toulouse, France\\[0pt]
\vspace*{0.25cm}
{\Large A. Ajabshirizadeh}\\[0pt]
\vspace*{0.25cm}
Research Institute in Astronomy, University of Tabriz, Tabriz, Iran\\[0pt]
\end{center}

\begin{abstract}

Irradiance variability has been monitored from space for more than two decades. Even if data are coming from different sources, it is well established that a temporal variability exists which can be set to as $\approx$ 0.1\%, in phase with the solar cycle. 
Today, one of the best explanation for such an irradiance variability is provided by the evolution of the solar surface magnetic fields. 
But if some 90 to 95\% can be reproduced, what would be the origin of the 10 to 5\% left? Non magnetic effects are conceivable. In this paper we will consider temporal variations of the diameter of the Sun as a possible contributor for the remaining part. Such an approach imposes strong constraints on the solar radius variability. We will show that over a solar cycle, variations of no more than 20 mas of amplitude can be considered. Such a variability --far from what is reported by observers conducting measurements by means of ground-based solar astrolabes-- may explain a little part of the irradiance changes not explained by magnetic features. 
Further requirements are needed that may help to reach a conclusion. Dedicated space missions are necessary (for example PICARD, GOLF-NG or SDO, scheduled for a launch around 2008); it is also proposed to reactivate SDS flights for such a purpose. 
\end{abstract}

\section{Introduction}

Total Solar Irradiance (TSI) has been measured from various spacecrafts platforms (and some rockets) since November 16, 1978. TSI refers to the total electromagnetic energy received by a unit area per unit of time, at the mean Sun-Earth distance (1 AU). Historically, TSI has been called the {\it ``solar constant"} by A. Pouillet in 1837. Since then, measurements have been made by a long series of observers as late as 1942 when Abbot
set the conventional value to 1.98 $\pm$ 0.05 cal$/$min$/$cm$^2$ (i.e. 1381 $\pm$ 35 W$/$m$^2$, close to the modern one, 1366 $\pm$ 1.3 W$/$m$^2$ (de Toma et al., 2004). At that time, it was recognized, notably by S.P. Langley, that the solar output might not be constant, by the order of $\pm$ 0.05$\%$. However, until to the 70's, a great number of investigators favored the notion of an unchanging Sun. Evidence suggesting variation of the solar output was thus very controversial. The difficulties in measuring the solar constant through the ever-changing Earth atmosphere precluded a resolution of this question prior to the space age.

\begin{figure}[t]
\begin{center}
\includegraphics[width=.7\textwidth]{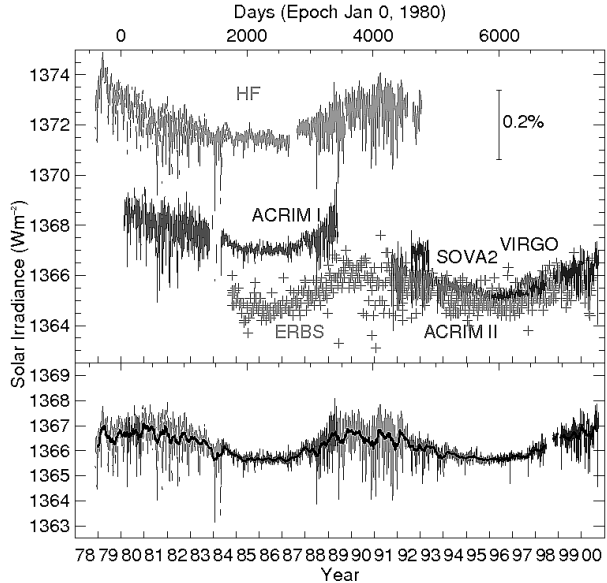}
\end{center}
\caption[]{The various total irradiance time series are presented
on the upper panel, the composite total solar irradiance
is shown on the lower panel (updated from Fr\"ohlich, 2000; courtesy J. Pap, 2003). }
\label{pap_eps1}
\end{figure}

Today there is irrefutable evidence for variations of the total solar irradiance. Measurements made during 26 years (two and an half solar cycles) show a variability on different time-scales, ranging from minutes up to decades, and likely to last even longer. Despite of the fact that collected data come from different instruments aboard different spacecrafts, it has been possible to construct a homogeneous composite TSI time series, filling the different gaps and adjusted to an initial reference scale. Fig. \ref{pap_eps1} shows such a composite, details of which can be found in Pap (2003). The most prominent discovery of these space-based TSI measurements is a 0.1$\%$ variability over the solar cycle, values being higher during phases of maximum activity  (Pap, 2003 and other references therein). Since the data do not come from a unique instrument, it is not yet clear if the minimum level of TSI varies from one cycle to another one.

The physical causes of these variations still remain a subject of debate. If we know that surface magnetism might be the primary cause, we are not sure that other factors would not have a role to play. Among them, one possible could be temporal variations of the radius of the Sun, others being mainly effective temperature and solar oscillations frequencies. We will deal here merely with radius changes. However even if the Sun's radius is the parameter implied in standard solar models, we prefer to speak in terms of the Sun's shape. In any case, the physical link between this geometrical shape and radiance is more direct than the simple correspondence between radius and radiance. We are from the start very careful on this question, and we will later explain why. 

Curiously, concerning the current situation of a possible temporal variation of the radius, the situation is the same as that which prevailed for the solar constant at the time of Abbot. The lack of consensus on the subject is mainly due to measurements made by means of ground-based instruments suggesting a temporal variation, but contaminated by atmospheric artefacts.
Several space missions to measure simultaneously the TSI and the shape of the Sun have been proposed, but so far, none has been launched yet.

Understanding and modeling TSI is important for at least three main reasons. First, for the reconstruction of the TSI before the satellite era. Second, for predictions, and third, linked to the others, for the radiative forcing on the Earth's troposphere. A successful model of TSI is therefore of particular interest both for {\it space weather} (subject of this volume) and for climatic impacts. 

Based on the outlines mentioned above, this paper is divided into three sections. In the first one, we will briefly recall our present understanding on TSI changes, emphasizing observed differences between irradiance and magnetic field variations. In the second section, we will describe our concept of the Sun's shape. Then we will show how observed irradiance changes put strong constraints on temporal radius variations. We conclude by recommanding  a satellite mission dedicated to measuring in real time the solar asphericity-luminosity parameter $w$, as described by Sofia (1979) or Lefebvre and Rozelot (2004). 

\section{Solar irradiance modeling}

The different time-scales found in the TSI data --very short-term variations (minutes, hours, some days), middle-term changes (months, years) and long-term changes (centuries, millenia)-- are linked with different physical mechanisms. While granulation and supergranulation may explain variations at the scale of the hour of the irradiance (Fr\"ohlich et al, 1997), $p$-modes 
variability may justify  variability at the scale of the minute (Woodward and Hudson, 1983). A number of correlative studies showed that sunspots, faculae and the magnetic network play a major role on short-term changes, albeit mechanisms that cause the 11-year cycle variability are still debated. Over longer periods of time, we do not really know what the causes of changes in TSI are. Only models that allow irradiance variations to be reconstructed on a large range of time-scales can provide a comprehensive answer to this open question (see also for more details, among several authors, Solanki and Fligge, 2002 or Pap, 2003).

Presently, the most successful models assume that surface magnetism is responsible for TSI changes on time-scales of days to years (Foukal, 1992, Solanki et al., 2005). For example, modeled TSI versus TSI measurements made by the VIRGO experiment aboard SOHO, between January 1996 and September 2001, show a correlation coefficient of 0.96 (Krivova et al., 2003).
However, significant variations in TSI remain unexplained after removing the effect of sunspots, faculae and the magnetic network (Pap et al., 2003). 
Moreover, there is a phase shift between TSI and magnetic variations at the beginning of solar cycles 22 and 23. In addition, solar cycle 23 is much weaker than the two previous cycles, in terms of magnetic strength, while TSI remains at about the same level (Pap, 2003). In the same way, de Toma et al. (2001) reported that TSI variations in solar cycle 23 implied additional mechanisms, other than surface magnetic features alone. Ermolli et al. (2003), from full disk images obtained with the Rome PSPT instrument argued that the network contribution to TSI cycle variations is significant. Finally, Kuhn (2004) described how the magnetic fields originating from the tachocline must increase the luminosity; the conclusion is that {\it ``sunspots and active region faculae do not explain the observed irradiance variations over the solar cycle"}. There is thus a need for a better identification of all mechanisms responsible for solar irradiance variations. Among them, the shape of the Sun must be explored.

\section{The shape of the Sun and its possible variability}

Solar semi-diameters\footnote{Due to the design of the instrument, measurements by means of solar astrolabes do not give a unique diameter but two radii forming between them a small angle $\epsilon$; hence it is preferable to speak in terms of semi-diameters.} has been measured for more than 20 years from several places in the world by means of solar ground-based astrolabes (France, Chile, Brazil and Turkey), and diameters since 2000, with the scanning heliometer of the Pic du Midi Observatory. Other Earth-based measurements are available, made by other techniques, mainly at Mount Wilson (USA) or Iza$\tilde{n}$a (Spain). High-altitude sporadic  
measurements were made during balloon flights carrying the Solar Disk Sextant (SDS) instrument (Sofia et al., 1994). Finally, a few measurements have been made from space by the MDI instrument aboard SOHO (Emilio et al., 2000).
The direct comparison of the temporal series acquired by astrolabes show discrepancies both in amplitude and in phase, albeit a more complete agreement can be found when sorting data by bins of heliographic latitudes (Lefebvre, 2003). 
In contrast, results obtained at the Pic du Midi Observatory (Rozelot et al., 2003) show a very strong similarity with those obtained at Mount Wilson (Lefebvre et al., 2004): the shape of the Sun departs from a perfect ellipsoid. The deviations (asphericities) which represent latitudinal variations of the solar radius, point out a bulge near the equator extending to 20-30 degrees, followed by a depression at higher latitude, near 60-70 degrees. Possible theoretical explanations for such a distorted shape $R$ = $f(\theta)$, ($\theta$ being the latitude) have been published in recent papers (see for instance Dikpati et al., 2001, and their Fig. 1, or Lefebvre and Rozelot, 2004). For our purpose, it is clear that the unperturbed value of solar irradiance ($I_0$) will differ from the one computed with a pure spherical approach ($R$ = constant = $R_{\odot}$). 

An alternative way of determining temporal variations of the solar radius is by heliosismology, and more precisely through $f$-modes of oscillations.
Estimates of solar radius variations deduced that way cannot be directly compared to visual observations. However, radius changes of an amplitude of $\Delta R$/$R$ = 2 $10^{-5}$ have been obtained by Schou (2004) over half a solar cycle (1997-2004). 

\section{Luminosity-radius constraints} 

The luminosity at the surface of the Sun is determined by the radius and the effective temperature ($L=\sigma T^4 S$, where $\sigma$ is the Stefan constant). The spheroidal envelope of surface $S$ depends on the equatorial and polar radii. Differentiating the surface over the radius $dR$ (assuming $dR_{eq}$ = $dR_{pol}$), one gets
a relation between $dI$ and $dR$. Adopted values of the equatorial and polar radii are 
$R_{eq}$ = 695991.756 km and $R_{pol}$ = 695984.386 km (Rozelot and Lefebvre, 2003).
Successive steps of $dR$ lead to corresponding $dS$ (given by the geometry involving $R_{eq}$ and $R_{pol}$)\footnote{The formula is rather complex but detailed computations can be found in Lalande, 1762, ``l'Astronomie", vol. de l'Acad\'emie des Sciences de Paris, reprinted by Johnson Corp., New York.} and $dL$ through the formula 
$dI = dS~ I_o/S_o$, where $I_o$ is the mean of the whole data set of the total solar irradiance. The final step consists of searching the best sinusoidal fit 
$ I_{calc}=I_o+ dI sin(2\pi t/P+\phi)$
where $t$ is the time variable, $\phi$ the best-fit phase and $P$ the best-fit period. 

\begin {figure}[t]
\begin {center}
\includegraphics[width=8.9cm,
angle=-90 ]{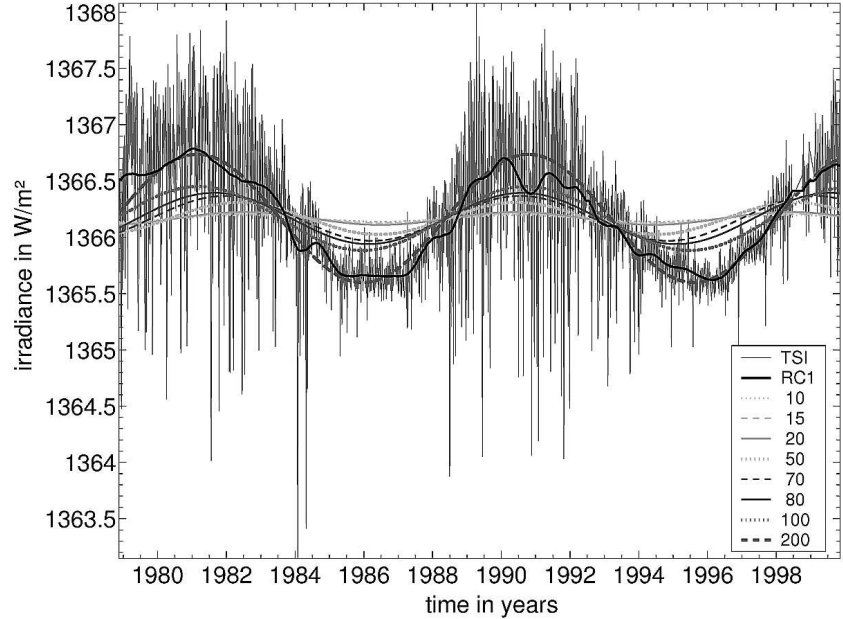}
\caption {Variation of solar luminosity associated with a temporal sinusoidal change of the solar radius. Dashed gray lines, solid grey line or dot-line represent the Total Solar Irradiance (TSI) and
the plain black curve its RC1 component (trend, see text). The successive sinusoidal curves show the irradiance variations computed through successive values of $dR$: the trend is rather well explained by a 200-mas radius amplitude variation, but such a large value is excluded. However up to a 10-mas radius amplitude variation still explains part of the trend. }
\label{fig:RL1}
\end {center}
\end {figure}

Four studies have been made according to the different solar cycles. The first one takes into account the whole irradiance data set (see Fig. \ref{fig:RL1}) and the three others take into account each separate solar cycle namely 21, 22 and 23.
These studies show that a variability $dR$ of 200 mas fits rather well the irradiance component RC1, that is to say the trend (the RCn components are the $n$th reconstruction in the decomposition by the Singular Spectrum Analysis).

In other words, a variation of $dR$ = 200 mas over a solar cycle may explain about 90 to 95\% of the trend, that is to say the same amount as solar surface magnetism alone! Such a huge solar radius variation is absolutely incompatible
with heliosismology (albeit such an amplitude is sometimes given by solar astrolabe observers). Nevertheless a $dR$ = 20 mas, of the order of magnitude of what is found from ground-based observations at the Pic du Midi, is still of a sufficient level to explain part of the trend. 
The dependence of the irradiance trend with the peak to peak temporal amplitude of the radius variations can be deduced from the median of the series obtained when dividing the trend (RC1) of the whole irradiance data by the best sinusoidal curve fit computed for each $dR$. Estimates of this median show that 
a {\it maximum} $dR=20$ mas amplitude could explain a {\it maximum} of 10\% of the trend. Such faint variations will teach us how the Sun's convective envelope responds to emergent energy fluctuations. They are detectable: the SOHO/MDI experiment has proved that exceedingly small solar shape fluctuations are measurable from outside our atmosphere (Kuhn et al., 1998). 

\section{Conclusion}

In this study we have shown that part of the TSI variations could be explained by variations in the solar radius (more precisely in the solar shape changes). Even if a modulation of 200 mas may explain the quasi-totality of the irradiance trend, such a large variability is already excluded. However, an amplitude of a maximum (peak to peak) 20 mas over one solar cycle explains 10\% of this trend, and 10 mas, $\approx$ 5\%. It is thus suggested that so far non explained variations of TSI could draw their origins from non-magnetic mechanisms. Such an approach has also been studied by Callebaut, Makarov and Tlatov (2002) and Li et al. (2003). The first authors developed a self-consistent model to explain cyclic variations of solar radius variations: the decrease in luminosity during half a solar cycle, balanced by gravitational energy, yields radius variations of $\approx$ 11 mas. The latter authors constructed models of the solar structure in which cyclic radius variations have been taken into account. 
At last, recent space-based data obtained with SOHO/MIDI by Kuhn et al. (2004) place limits on solar radius variations, $dR < 7$ mas, and $w < 0.1$, still leaving room for such changes.

While a considerable amount of data has been gathered on TSI variations since 1978, we still significant lack of information on the possible temporal changes of the solar diameter (and of the Sun's shape). In-flight missions are necessary, particularly dedicated missions which could measure in real time 
and simultaneously, irradiance and the  asphericity-luminosity parameter $w$, (for instance aboard PICARD, GOLF-NG or the next SOHO satellite, called SDO). Until such missions take place, though not scheduled before 2008, we propose to reactivate annual SDS balloon flights based upon a joint European-Aerican program (Rozelot and Sofia, 2004). 

Finally, we would like to emphasize the need for a better understanding of TSI variability to discern its role in space-weather and its impact on the terrestrial climate. 

\vspace{0.5cm}
\noindent
\Large{\bf{Acknowledgements}}\\

\noindent
\small{The authors wish to thank P. Scherrer for useful suggestions on the manuscript.}

\end{document}